\documentclass[prd,amsfonts,nofootinbib,aps]{revtex4}

\begin{document} 

\title{Quantum mechanics, strong emergence and ontological
non-reducibility}

\author{Rodolfo Gambini$^1$ Luc\'{\i}a Lewowicz$^2$ Jorge
  Pullin$^3$\footnote{Corresponding author. Email:
    pullin@lsu.edu. Tel/Fax: +1 225 578 0464}}
\affiliation{1. Instituto de F\'{\i}sica, Facultad de
  Ciencias, Universidad de la Rep\'ublica, Montevideo, Uruguay \\
2. Departamento de Historia y Filosof\'{\i}a de la Ciencia,
Universidad de la Rep\'ublica, Montevideo, Uruguay \\
3. Department of Physics and Astronomy, Louisiana
State University, Baton Rouge, LA 70803 }

\date{August 5th 2013}

\begin{abstract}
  We show that a new interpretation of quantum mechanics, in which the
  notion of event is defined without reference to measurement or
  observers, allows to construct a quantum general ontology based on
  systems, states and events.  Unlike the Copenhagen interpretation,
  it does not resort to elements of a classical ontology. The quantum
  ontology in turn allows us to recognize that a typical behavior of
  quantum systems exhibits strong emergence and ontological
  non-reducibility.  Such phenomena are not exceptional but natural,
  and are rooted in the basic mathematical structure of quantum
  mechanics.

{\bf Keywords:} Interpretation of quantum mechanics;  emergence; downward
causation. 
\end{abstract}

\maketitle

\section{Introduction}

There has been a recent proposal to solve the {\em problem of time} in
background independent systems like general relativity (Gambini {\it
  et al.} 2009). In it, one has to take into account how quantum and
gravitational effects influence the measurement of time. The latter
ceases to be an ideal variable and becomes a physical variable subject
to uncertainties of measurement and fluctuations of fundamental and
inescapable nature (Frenkel 2010 and references therein; Ng and Van
Dam 1993). The description of the evolution of
quantum physical systems in terms of physical time changes the
traditional Schr\"odinger picture (Gambini {\it et al.}
2007). Although the departures from the traditional theory are very
small, they are enough ---as we will argue later--- to lead to an
interpretation of quantum mechanics that solves the measurement
problem in purely quantum terms. Such interpretation is known as the
Montevideo interpretation (Gambini {\it et al.} 2011). A philosophical
assessment of the interpretation was recently given by Butterfield
(2014). In this paper we would like to argue that having at hand a
realist interpretation of quantum mechanics allows to establish a
general ontology of states and events that challenges some ontological
prejudices inspired by classical systems that have clouded the
understanding of the problem of emergence.

The organization of this paper is as follows. In section II we briefly
review some notions of quantum mechanics. In Section III we review the
measurement problem. Section IV introduces a new interpretation of
quantum mechanics. In section V we discuss how to construct a new
general ontology based on the new interpretation. Section VI discusses how
emergence arises naturally from the new ontology. We end with a
summary. 

\section{Quantum mechanics}

Quantum physics has two fundamental properties from which very
important consequences follow: it is quantum and it is
probabilistic. First and foremost, it is quantum; the very name of
this theory has to do with the fact that many fundamental quantities
do not take continuous values, but certain preselected discrete
values.  For instance, in the quantum theory the orbits of the atom
can only have certain values for the energy. To this fundamental
property we owe the existence of matter organized in atoms, molecules
and solid bodies.  Every element or substance has a discrete set of
behaviors that completely characterize it.  Secondly, quantum physics
is probabilistic. A complete knowledge of the state of a system will
not allow us in most cases to know with certainty its future behavior.
We have a great body of evidence indicating that the ignorance behind
the probabilities is fundamental in nature (Schlosshauer {\it et al.}
2013). A more complete knowledge of the system is impossible: for
instance, we just cannot predict when a given radioactive atom will
emit a particle.

Many times there is the tendency to think that quantum phenomena are
only relevant at microscopic scales and that classical physics is
enough to describe phenomena at ordinary scales. Such belief is
incorrect, quantum physics underlies many fundamental physical
phenomena at ordinary scales: the stability of atomic and molecular
structures, the very existence of solid matter and its electric,
thermal, and optical properties, the origin of chemical reactions, the
colors of certain substances are just some examples that rely
crucially on quantum mechanics for their existence and explanation.

In spite of their relevance for macroscopic phenomena, the quantum
processes underlying physics at ordinary scales are quite far from the
ordinary experience. For instance in the well known double slit
experiment, quantum systems like electrons neither behave like waves
nor like particles. They have a dual behaviour. Each ordinary electron
produces a dot on the screen as it a classical particle would, but
when a large number of electrons have gone through the slits they
reproduce the interference pattern of a wave.  Even if the electrons
go through the slits one by one they reproduce the wave pattern.  One
can say that the electron behaves as a particle when it is detected by
the screen and as a wave when it travels undetected through the slits,
interfering with itself.  The double slit experiment allows us to
introduce some concepts that will be useful in the forthcoming
discussion of the measurement problem and the new interpretation of
quantum mechanics. The behaviour of an electron in the photographic
plate suggests that the electron has a certain {\em disposition} to
produce an {\em event}, the appearance of dot in a given region
of the plate with a given probability. This disposition for certain
behaviors is determined by the wavefunction or {\em state} of the electron.
While it propagates without being observed, the state given by the
wavefunction obeys a wave equation: the Schr\"odinger equation.
In quantum mechanics {\em measurements} are events to which we assign
quantitative properties. For instance, in the double slit experiment,
the position of the dot on the photographic plate.

\section{The measurement problem}

In order to explain the problem of measurement in quantum physics, let
us come back to the previously mentioned double slit experiment. We
have seen that electrons traversing one by one the double slit seem to
interfere with themselves. One may wonder what happens if one
illuminates one of the slits in order to determine whether the
electron goes thorough that slit or the other.  It happens that while
the electrons are not illuminated as they travel from the source to
the plate, they interfere with themselves and produce an interference
pattern. The state of each electron may be described by a
superposition corresponding to electrons going through both slits.
When the electrons are illuminated when they pass through the slits,
allowing to determine which of them they pass through, the state of
the system suffers a sudden change and the electrons are forced to
choose through which slit they will pass. What we observe in this case
is the pattern produced by classical particles with two bright regions
in front of each slit on the photographic plate.  The state of the
electron after being illuminated and before arriving to the plate is
now given by a new kind of state known as statistical mixture. With
certain probability, the electron is in a state localized in front of
the upper slit and with the complementary probability will be in a
state localized in front of the lower slit.

The passage from a superposition to a statistical mixture cannot be
explained by the Schr\"odinger equation. Quantum mechanics therefore
seems to require two different, incompatible, types of evolution: an
evolution described by the Schr\"odinger equation and a different
evolution that occurs when the system is measured, and an event is
produced.  But in a quantum Universe as the one we seem to live in,
the measurement process should be analyzed entirely in quantum
mechanical terms. What could distinguish a measurement from any other
physical process to endow it with a different type of evolution?

The measurement problem can be summarized as follows: If one
attempts to describe the measurement process using the usual evolution
equation of quantum mechanics, that is the Schr\"odinger equation, one
obtains that, after a measurement, the measuring device is left in a
superposition in which the needle of its gauge is not pointing in a
definite direction, but is generically pointing in all possible
directions simultaneously. This result is not what happens when one
performs an actual experiment; there the needle points to a definite
position.

\section{A new realist Interpretation of quantum mechanics}

The founding fathers of quantum mechanics, Bohr and Heisenberg
basically attempted to solve the measurement problem by denying the
possibility of describing the world in pure quantum mechanical terms
(Bohr 1995, Heisenberg 1958/2007). The basic problem with this
proposal is that it needs to assume the existence of two different
ontological realms: the classical ---macroscopic--- and the quantum
---microscopic---, and the analysis of their interaction is outside
the scope of any available theory.  Moreover, in the last decade
physicists have discovered in increasing numbers macroscopic systems
whose description requires quantum mechanics, and many became
convinced that the world has a fundamental quantum nature
(Schlosshauer {\it et al.} 2013). Even Bohr used quantum mechanics for
the description of the measuring device to address Einstein's
objections (Bohr 1949). The closest thing we have to an explanation
for the measurement process within the quantum theory is environmental
decoherence (Schlosshauer 2010). It is based on the fact that when
the state of a quantum system interacts with an environment with an
enormous number of (microscopic) degrees of freedom, the quantum
system suffers transitions that almost look like the abrupt evolutions
one needs to postulate in measurements.

Even if they are difficult to detect, quantum superpositions
are still there and may be in principle observed. They simply are
distributed in the environment and recovering them would require the
observation of a system with a large number of degrees of freedom.
Environmental decoherence is important because the measurement
processes involves interactions with macroscopic systems with many
degrees of freedom. Interactions with the environment were neglected
for decades and the relevance of this effect was only recognized in
the 1980's.  

There is not universal agreement that decoherence solves the
measurement problem. To begin with, as we just mentioned, quantum
coherence is still there after an interaction with the environment and
may be recovered. This is not what is supposed to happen in a
measurement in quantum mechanics. Moreover, since the evolution of the
system plus measuring device plus environment is unitary, there is no
sense in which a measurement happens at some instant in time. And
there is no objective criterion for when an event results from the
measurement. Decoherence yields quantum states that resemble those of
a system after a quantum collapse takes place, but there is no guide
on what ``resemble'' means. Moreover there is no guide to say when
the quantum state transitions from a superposition of coexisting
possible outcomes to a set of exclusive outcomes with respective
probabilities.

A new interpretation of quantum mechanics was recently proposed,
called the Montevideo Interpretation (for a recent pedagogical review
see Gambini and Pullin 2015). It is based on supplementing
environmental decoherence with a more careful investigation of the
measurement process, in particular taking into account the limitations
of measurement that gravity imposes. It has the novelty of the
inclusion in the quantum description of another factor up to now
neglected. In the standard Schr\"odinger description of the evolution,
time is treated as a classical external parameter, but time is
actually measured by physical clocks that obey quantum mechanical
laws. Quantum measurements of time have a limited precision. This
limitation arises from quantum fluctuations and gravitational time
delay effects (Frenkel 2010).  One can show that a quantum mechanical
treatment of time combined with fundamental limitations of
measurements stemming from general relativity, leads to a modified
Schr\"odinger evolution that allows transitions between quantum
superpositions and statistical mixtures (Gambini {\it et al.} 2007)
For a macroscopic object also subject to environmental decoherence
this transitions may occur quite rapidly.

It is worthwhile remarking here that this interpretation results from
a previous study and solution (Gambini {\it et al.} 2009) of the issue
of time in quantum generally covariant theories. In these theories,
like in any version of quantum gravity, the only well defined
quantities are constants of the motion, in fact invariants under
general coordinate transformations. It was shown that it is possible
to describe the evolution using purely relational properties of
quantum Dirac observables of which one of them plays the role of a clock.

When one takes into account the limitations in measurement imposed by
quantum mechanics and gravity, the states that decoherence produces
are indistinguishable from those produced by the measurement
postulate.  The ``almost'' of the standard approach to decoherence is
removed by fundamental limitations of the theory itself and the
transitions from superpositions to statistical mixtures required to
explain measurements are a consequence of the theory. This in turn
supplies an objective criterion that says when and what events may
occur. Events occur when the state of a system resulting from a full
quantum mechanical evolution becomes indistinguishable from a
statistical mixture.

Events arise as random choices of the system when this criterion
---the appearance of a statistical mixture--- is fulfilled. It is
important to remark that for quantum systems interacting with a
macroscopic environment that has many degrees of freedom, events will
be plentiful. They not only occur on measuring devices, they occur
around us all the time.  Measurements are nothing else but the assignment
of quantitative properties to events occurring in measuring devices.
In light of this new interpretation there is nothing mysterious or
exceptional in the measurement process.  

The philosopher of physics Jeremy Butterfield has recently assessed
the significance of this interpretation, known as the Montevideo
Interpretation and its analysis of time, and suggested that the
resulting mechanism of decoherence is also compatible with a Many
World interpretation (Butterfield 2014).

\section{A quantum ontology}

It is important to remark that having a realist interpretation of
quantum mechanics not only allows us to understand the measurement
process; it also allows understanding how a world with uniquely
defined properties arises from a quantum mechanical world of
potentialities\footnote{This not only applies to the Montevideo
  interpretation but may apply to other realist interpretations like
  the modal ones, although this has yet to be studied.} . This leads
us to a new ontology based on quantum mechanics that we shall briefly
discuss.  A general quantum ontology addresses the question of which
fundamental entities exist and their characteristics. Traditional
ontologies are deeply ingrained in classical physics. For instance the
concept of object as bundle of properties may sound plausible in
classical physics but it is clearly superseded by quantum physics
where, typically, systems do not have properties until events are
produced.  

To have a quantum theory with an interpretation is equivalent to
knowing what beliefs about the world are allowed by the formalism. For
this purpose, one relates the elements of the formalism with a
consistent ontology.  The interpretation here considered makes
reference to primitive concepts like system, state, events and the
properties that characterize them. Although these concepts are not new
and are usually considered in quantum mechanics, one can assign them a
univocal meaning only if one has an interpretation of the theory.  For
example, events could not be used as the basis of a realistic ontology
without a general criterion for the production of events that is
independent of measurements. Moreover, the concepts of state and
system only acquire ontological value when the events also have
acquired it.

Based on this ontology, objects and events can be considered the
building blocks of reality.  Objects will be represented in the
quantum formalism by systems in certain states.  In the new
interpretation, events are the actual entities. On the other hand,
states describe the potentialities or dispositions of the systems for
the production of certain events.  Concrete reality accessible to our
senses is constituted by events localized in space-time.  As Whitehead
(Whitehead 1925/1997) recognized: ``the event is the ultimate unit of
natural occurrence.''

Events come with associated properties. Events and properties in the
quantum theory are represented by mathematical entities called
projectors. Quantum mechanics provides probabilities for the
occurrence of events and their properties.  When an event happens,
like in the case of the dot on the photographic plate in the
double slit experiment, typically many properties are actualized. For
instance, the dot may be darker on one side than the other, or may
have one of many possible shapes.  

Systems in a given state are what
we colloquially call ``objects''. All hydrogen atoms are identical
systems from the point of view of their potential behaviors, but their
specific behavior at a given instant is determined by which state the
atom is in. The quantum formalism associates to each system a Hilbert
space, and to each state a vector or more generally a trace one
positive definite self-adjoint operator.  For instance the hydrogen
atom system could be represented by the vector space of all its possible
pure states. A basis from this space could be given by the standard
orbitals associated with its stationary states. One could identify the
hydrogen system with the element hydrogen. On the other hand, a
particular hydrogen atom is a hydrogen system in a given state, it is
an example of what we call an individual object and it has a precise
disposition to produce events. 

Notice that this quantum ontology
has deep differences with the kind of ontologies that are usually
considered in the discussion of the issue of emergence.  For instance
as noticed by McLaughlin, Brian and Bennett, Karen, ``Supervenience'',
The Stanford Encyclopedia of Philosophy (Winter 2011 Edition), Edward
N. Zalta (ed.) ``Supervenience is a central notion in analytic
philosophy. It has been invoked in almost every corner of the
field. For example, it has been claimed that aesthetic, moral, and
mental properties supervene upon physical properties. It has also been
claimed that modal truths supervene on non-modal ones, and that
general truths supervene on particular truths. Further, supervenience
has been used to distinguish various kinds of internalism and
externalism, and to test claims of reducibility and conceptual
analysis.'' In the same article supervenience is defined in the
following manner: A set of
properties A supervenes upon another set B just in case no two things
can differ with respect to A-properties without also differing with
respect to their B-properties. In slogan form, ``there cannot be an
A-difference without a B-difference''.  As we shall see in the
forthcoming sections this definition is particularly ill suited for
the quantum description of the world where a system may not even have
any property. Properties typically belong to events. Some properties
may be ascribed to states only in some very specific situations.

\section{Emergence}

Emergent phenomena are said to arise out of and be sustained by more
basic phenomena, while at the same time exerting a ``top-down''
control, constraint, or some other sort of influence upon those very
sustaining processes.  We are considering here the most stringent
conception of emergence, which Mark Bedau (2003) calls
strong emergence which assumes ontological novelty and ``irreducible
causal powers''. He considers that ``All the evidence today suggests
that strong emergence is scientifically irrelevant... There is no
evidence that strong emergence plays any role in contemporary
science. The scientific irrelevance of strong emergence is easy to
understand, given that strong emergent causal powers must be brute
natural phenomena. Even if there were such causal powers, they could
at best play a primitive role in science. Strong emergence starts
where scientific explanation ends.'' This position is not exceptional
but rather the rule among many contemporary philosophers.

In our view, efforts to explain this type of emergence have failed
mainly because they assume implicitly an ontology inspired in
classical physics.  We will show in this paper that certain quantum
mechanical systems present precisely this kind of top-down control.
We will attempt to show that quantum mechanical systems have
ontologically new properties and downward causation where
macro systems have effects on their micro components.

\subsection{Ontologically new properties}
 
Let us start by showing that quantum systems may have ontologically
new properties. This has already been noted by Howard (2007), whom
however, lacking an interpretation of quantum mechanics can only
recover partially the results we obtain. In particular, here we can
discuss the properties of a quantum object, whereas in his case the
emergent properties are only about measurements, and downward
causation is not discussed.

Quantum systems may be in certain quantum states,
called entangled, that have well defined properties that neither
follow from the properties of parts, nor from relations among them.
To understand this statement better let us review how entangled states
are defined and contrast them with systems in classical states.
In classical physics the state of a system of particles is simply the
union of the states of each of the particles given by their positions
and velocities at a given instant. Its knowledge determine all the
properties of the system. For instance the energy.  All the properties
of a classical system are functions of the properties of its
components.  In quantum mechanics things are very different.  Most of
the properties of a system do not have well defined values until
measured; for instance the position of an electron in the double slit
experiment is not well defined until a dot is produced in the
photographic plate and it is detected.  

In spite of the fact that in general one cannot attribute properties
to states, quantum systems in a pure state have some well defined
properties. In order to exemplify this, let us consider a spinning
particle like the electron. While a classical particle rotating along
an internal axis may have a continuum of values for the projection of
its angular momentum on a given direction of space, in quantum
mechanics its projections along an arbitrary axis are quantized. For
instance, a spinning particle like the electron can only take two
possible values of its projection along an axis z: up or down.  Given
a spinning particle like the electron one can measure its spin
projection along z by using a Stern Gerlach device. It mainly consist
in a magnet with S-N poles oriented along the axis in question, z in
this case, and a photographic plate.

When one performs repeated measurements on a particles in a generic
state one observes dots appearing in the upper region with certain
probability an in the lower with the complementary probability.  When
the electron is in a state that leads with certainty to a dot in the
upper region, one may say that it is in the state $\vert z up
\rangle$. In this case one may assign the property ``z up'' to the
state. This is the only property that one can assign to this
state. The measurement of any other projection of the spin along a
different direction will not lead to a unique value: i.e. always up or
always down.  {\em It is only when one knows with certainty what will be
the behavior of the system in certain state that one may assign a
property to the state}. Recall that while events always have many
properties, and are completely characterized by them, one can only
assign a property to a state if one knows with certainty to which
event it will lead in certain measurement.  Systems composed of
several particles may also have states with some properties with
well defined values. However, these properties may refer to the system
as a whole and, in these systems, there may not be any property for
the states of individual particles with well defined values. These
composite systems are examples of what is known as systems in 
entangled states.

More in general, systems in entangled states are those that have
properties with well defined values than cannot be inferred from those
of their constituent parts.  As we mentioned, and will show in the
following example, it might even be the case that the states of the
constituent parts have no well defined properties and yet the state of
the whole system does.  

Consider two electrons with spin in the z direction in a state
\begin{equation}
  \vert \psi_0\rangle= \frac{1}{\sqrt{2}} 
\vert 1,z,{\rm up}\rangle \vert 2,z,{\rm down}\rangle
+\frac{1}{\sqrt{2}} \vert 1,z,{\rm down}\rangle
\vert 2,z,{\rm up}\rangle.
\end{equation}
This would represent a superposition of two electrons in different
positions 1 and 2 with opposite spins. Neither the state of particle 1
nor the one of particle 2 have well defined properties.  No matter
what projection of the spin one measures, one has a probability one
half of measuring up and one half of obtaining down.  Even though each
of the entangled electron do not have well defined properties for
their spin components, the total system does. For instance, one can show
that it has total spin $s=1$ in Planck units, and z component of the
total spin $s_z =0$. It is only when the observations made on particle
1 and 2 are compared that one can discover the properties of the total
system. One could also determine these properties if the complete
system is measured.  The constituents therefore now form an
inseparable unit whose state is endowed with properties without the
states of the individual particles having any property ---any spin
component--- with well defined value.  Notice that we made no
statement about where the spinning particles are located in space.

This holistic behavior is actually not an exception but is the generic
behavior of two quantum systems after an interaction.  For instance,
the precise vibrational modes of a molecule depend on the entangled
system of electrons and nuclei.  The emergent properties of such
systems are crucial for explaining (at least some) chemical or
biological properties in terms of physics. For instance, the
vibrational modes of molecules are key to explaining why water is
transparent.  Ontological novelty manifests itself in the emergence of
new properties that do not result from properties of the parts. They
arise only when the composite structure is constituted.  The
philosopher of science Paul Teller (1986) was the first in
noticing that quantum phenomena show relations that do not stem from
non-relational properties of their relata, as is characteristic of the
classical description of the world. However, the ontological meaning
of his observation could not be elucidated because he could not
discard a purely instrumentalist interpretation and his use of the
notion of property is still reminiscent of classical physics.
Entangled systems present what Teller calls: relational holism.

\subsection{Downward causation}

A strong form of emergence also requires downward causation, namely,
the emergence of novel causal properties. Here a double goal arises:
to characterize such form of causality in physical terms and to show
that at least certain quantum systems, exhibit downward causation.  A
notion of causality that is suitable to the ontology of states and
events has been developed by Chakravartty (2007). He says ``So what
does it mean to say that causal properties 'do the work' of
causation?.... a causal property is one that confers dispositions on
the objects that have them to behave in certain ways when in the
presence or absence of other objects with causal properties of their
own.''  This dispositional idea of causality is the one we have
adopted in this work: recall that quantum states characterize
dispositions to produce events.  A system will present downward
causation if the parts have some behaviors that are dictated by the
state of the whole and that cannot be predicted from the knowledge of
the state of the parts.  The previous example of an entangled state
shows that in quantum mechanics there is state non-separability. The
states of the parts are just a statistical mixture of up components
and down components while the complete entangled state has more
information.  It is this state non-separability that leads to downward
causation. Let us study this behavior in some detail.
 
Let us assume George prepares the entangled state we discussed before
and sends particle 1 to Alice and particle 2 to Bob. We will assume
they both have devices that allow them to measure projections of spin
along x or y. Without any communication, Alice and Bob choose
independently and in a random fashion to measure one of these
projections every time they receive a particle. As we mentioned, if
the system is in this state it does not matter which component they
measure, they will have 50\% probability of measuring ``up'' and 50\%
``down''.  However, if they compare notes, they would realize that the
sequences will be correlated: whenever they happened to measure the
spin in the same direction their results were opposite, up for Alice
and down for Bob or vice versa. They could never have figured that out
by looking at the individual systems in isolation
The complete system has a certain non locality such that when one
electron chose to answer ``up'', the other necessarily needs to chose
``down''. Such correlation does not involve time, it is instantaneous.

There is a theorem, known as Bell's theorem (Mermin 1985), that establishes
that it is not possible to explain this kind of behavior assuming that
each part follows a pre-established set of instructions, in other
words, assuming that each part has some local hidden information
telling it how to act before each measurement.  Certain behaviors of
portions of a quantum system cannot be explained in terms of the
states of its components. To put it differently, the state of the
total system is not mathematically determined by the states of its
components.  Since we have characterized the causal power of a system
by its disposition to produce certain effects, downward causation will
be related to the non-separability of the quantum state of a whole and
the dispositions it induces on its components to produce events.

\subsection{A chemical example}

Let us consider a specific example relevant to chemistry: the
emergence of a molecule via the chemical bonding of its atomic
components. Let us take a simple molecule of ionic hydrogen, formed by
two atoms of hydrogen and stripping them of one electron. One can
actually show rather straightforwardly that a hydrogen atom and a
proton will tend to form a molecule. Using quantum mechanics one
computes the minimum energy that the system has when the electron
orbits around one of the protons and the minimum energy when it orbits
around both. It turns out that the latter is lower than the former
(Cohen-Tannoudji et al. (1978). As a consequence, after the molecule is formed
one would have to add energy (or apply a force) to separate it. A
chemical bond is therefore formed. This would be an example of how the
molecular structure of matter can be explained in terms of quantum
mechanics. Different molecules will require different studies, but in
all cases the bonded configuration will be the one with the lowest
energy. In complex molecules there can be several configurations that
correspond to local minima of the energy function and all of them will
be possible states of the molecule. Chemically, the bonds can
originate in different mechanisms depending on the molecule in
question, but they can all be explained in physical
terms. 

Ontologically this example shows how molecular structures emerge from
atomic ones. Chemical compounds arise as emergent phenomena of quantum
states where the subsystems lose individuality.  Epistemologically the
example illustrates how some chemical concepts can be reexpressed in
physical terms.  The quantum state of the molecule consisting of two
protons and an electron is an entangled state of the three
particles. Quite a few of the behaviors of the molecule can only be
explained with a state for the complete system, which cannot be
described just in terms of states of the parts. An example is given by
the vibrational modes associated with the oscillation of the protons
around their equilibrium positions. These have very real consequences:
they influence the ability of the molecule to emit or absorb light of
certain frequencies. The emission or absorption is directly related
with a rearrangement of the protons and electrons that are a
manifestation of the downward causation of the complete system on its
component parts. If one tried to describe that behaviour only in terms
of the component states, lets say of the electrons assuming fixed
positions of the protons, one could not reproduce the 
observed spectral lines, which are studied with great precision
experimentally. Entanglement in quantum chemistry has actually been
extensively analyzed, see for instance Huang, Wang and Kais (2006) and
references therein.

Entanglement is about correlations. In quantum chemistry calculations,
the correlation energy is defined as the energy error of the
Hartree-Fock approximation to the wavefunctions. Electron correlations are
important in many atomic, molecular and solid state properties.

Summarizing, in the language of the new ontology, for the above
example the {\em object} is the system composed of the two protons and
the electron in one of the system's possible states. An {\em event}
corresponds to the observation of a photon of a certain frequency when
it traverses a medium composed by the above objects. {\em Downward
  causation} occurs because one needs to consider the complete
entangled system to predict precisely the vibrational modes and from
them the optical behavior of the system (for instance absorption
lines).

\section{Conclusions}

Summarizing, by introducing in the analysis the use of quantum
physical clocks for the description of the evolution, one can show
that the standard Schr\"odinger equation needs to be slightly modified.
In certain conditions as the ones present in measuring devices one can
show that quantum superpositions may evolve into statistical
mixtures. This provides an objective criterion that says when and what
events may occur.  In terms of the notions of this new realistic
interpretation one can define a quantum ontology that leads to a
revision of the notion of matter and its potentialities.  This allows
us to notice that systems of particles in entangled states have new
capabilities and emergent properties.  The simplistic notion of
hierarchical systems with progressive complexity in which one goes
from subatomic particles to atoms to molecules or from cells to
tissues, organs, and living organisms, is inadequate.  The quantum
theory implies that the lower levels are modified even up to the point
where they lose part of their individuality when they integrate into
an entangled system in a higher level of the hierarchy. The emergent
structure has novel properties and downward causation.

The present analysis may be considered as epistemologically
reductionist because it allows to explain the appearance of novel
properties and downward causation in purely physical terms. However
ontologically it shows how different levels of reality may present
emergent new properties and states with top down causation. In that
sense, organization in higher level wholes are significant and the
efficacy of the higher levels undeniable: one has ontological
non-reducibility.

This work was supported in part by grant NSF-PHY-0968871, funds of the
Hearne Institute for Theoretical Physics, CCT-LSU and Pedeciba.

\parindent=0pt
\eject
{\bf References:}

\parskip 10pt
Bedau, M.: Downward causation and autonomy in weak
emergence. Principia, 6, 5 (2003) also Bedau M. and Humphreys P., (eds): Emergence: contemporary readings in philosophy and science. Cambridge: MIT Press (2008)

Bohr, N.: Discussions with Einstein on Epistemological Problems
in Atomic Physics. In Schilpp, P.: Albert Einstein: philosopher
scientist. Cambridge University Press, Cambridge, UK (1949) 

Bohr, N.: The Philosophical Writings of Niels Bohr, Vol. 3: Essays
1958-1962 on Atomic Physics and Human Knowledge. Ox Bow Press,
Woodbridge, CT. (1995)

Butterfield, J.: Assessing the Montevideo interpretation of quantum
mechanics. Stud. Hist. Philos. Mod. Phys. 10.1016/j.shpsb.2014.04.001. arXiv:1406.4351 (2014).

Cohen-Tannoudji, C., Diu, B., Laloe, F.: Quantum Mechanics. Wiley, New
York, NY (1991).

Frenkel, A.: A review of derivations of the space-time foam formulas
arXiv:1011.1833 (2010) and references therein. 

Gambini, R., Garc\'{\i}a-Pintos, L. and Pullin, J.: An axiomatic
formulation of the Montevideo interpretation of quantum
mechanics. Stud.Hist.Philos.Mod.Phys. 42, 256-263  (2011). 

Gambini, R., Porto, R. and Pullin, J.: Fundamental decoherence from
quantum gravity: A Pedagogical review. Gen. Rel. Grav. 39, 1143-1156 (2007).

Gambini, R., Porto, R., Pullin, J. and Torterolo, S.: Conditional
probabilities with Dirac observables and the problem of time in
quantum gravity. Phys. Rev. D79, 041501 (2009).

Gambini, R., Pullin, J.: The Montevideo Interpretaion of Quantum 
Mechanics: A Short Review. [arXiv:1502.03410 [quan-ph]] (2015).

Heisenberg, W.:Physics and Philosophy: The Revolution in
Modern Science. HarperCollins, New York  (1958/2007)

Howard, D.: In Murphy, N.  and Stoeger, W. (eds.), Evolution and
Emergence: Systems, Organisms, Persons. Oxford University
Press. 141-157 (2007)

Huang, Z., Wang, H. and Kais, S.: Journal of Modern Optics
53, 10 (2006)

Mermin, D.: Is the moon there when nobody looks?
Reality and the quantum theory. Physics Today  38 (April 1985) 

Ng, Y., Van Dam, H.: Limits to space-time
measurement. Mod. Phys. Lett. A9, 335 (1994)

Schlosshauer, M.: Decoherence: and the Quantum-To-Classical
Transition (The Frontiers Collection). Springer, Berlin (2010) 

Schlosshauer, M., Kofler, J. and Zeilinger, A. A Snapshot of
Foundational Attitudes Toward Quantum
Mechanics. Stud. Hist. Phil. Mod. Phys. 44, 222-230 (2013).

Teller, P: Relational Holism and Quantum Mechanics. Brit. J. Phil.
Sci. 37, 71 (1986) 

Whitehead, A. N.: Science and the modern world. Free Press,
New York (1925/1997)

\end{document}